\documentclass{article}
\usepackage{setspace}
\usepackage{graphicx}
\usepackage{amsmath}
\usepackage{float}
\usepackage{hyperref}
\usepackage{cleveref}

\newcommand{\1}{$\langle100\rangle$}
\newcommand{\2}{$\langle110\rangle$}
\newcommand{\3}{$\langle111\rangle$}

\title{Graph Theory Based Approach to Characterize Self Interstitial Defect Morphology}
\author{Utkarsh Bhardwaj$^a$ \and Andrea E. Sand$^b$ \and M. Warrier$^{a c}$}

\date{\small{%
    $^a$ Computational Analysis Division, BARC, Vizag, AP, India - 530 012\\%
    $^b$ Department of Physics, P.O. Box 43, FI-00014 University of Helsinki, Finland\\
    $^c$ Homi Bhabha National Institute, Mumbai, Maharashtra, India - 400 094
  }
}

\begin{document}
\maketitle
\begin{abstract}
The defect morphology is an essential aspect of the evolution of crystals' microstructure and its response to stress. Existing methods either only report defect concentration or characterize only some of the defect morphologies. The need for an efficient and comprehensive algorithm to study defects is becoming more evident with the increase in the amount of simulation data and improvements in data-driven algorithms.

We present a method to characterize a defect's morphology precisely by reducing the problem into graph theoretical concepts of finding connected components and cycles. The algorithm can identify the different homogenous components within a defect cluster having mixed morphology. We apply the method to classify morphologies of over a thousand point defect clusters formed in high energy W collision cascades. We highlight our method's comparative advantage for its completeness, computational speed, and quantitative details.
\end{abstract}



Defects in materials can have different morphology and sizes. The structural details of a defect decide its thermal stability, migration properties, and interactions with other defects \cite{BACON20001, SINGH1997107, OSETSKY200065, BECQUART200639, OSETSKY2002852}. These properties govern the microstructural evolution of the crystal and its mechanical and physical properties. For this reason, the morphology of defects has been of wide interest in the study of the effects of irradiation on materials and designing materials with desired properties.

The earlier simulations of defects in crystals were limited to lower primary knock-on atom (PKA) energies and fewer statistics due to limitations in computational power. The main focus of the results was defect concentration and cluster size distribution. The defect morphology studies were limited to studying particular defects of interest \cite{GAO2000213}. For the small datasets of clusters, it is possible to inspect the defect morphology visually. With continuous improvement in computational power, it is now possible to simulate bigger system sizes for longer durations, with more statistics. The promising applications of data-driven methods in various scientific fields have further garnered broad interest to create databases of atomistic simulations of materials such as DefectDB and CascadesDB \cite{CascadesDB}. This opens avenues for new data-driven, statistical findings if accompanied by efficient automatic tools that can reliably extract defect concentration and morphological details. 

The traditional methods for finding point defects, such as Wigner-Seitz (W-S) \cite{gibson1960dynamics,
nordlund1997point} and effective-sphere for finding displaced atoms (ES) \cite{Stoller2012293, warrier2015statistical}, do not differentiate the point defect clusters based on morphology. Dislocation loops have been identified in a number of ways, with the DXA algorithm [12] lately establishing itself as a useful tool in the field. However, it does not specify the morphology of non-dislocation defects that includes all small point defect clusters. Moreover, the fastest algorithms available for dislocation analysis can be memory intensive and slow as the system size grows. The traditional geometrical methods, such as common neighbor analysis, centrosymmetric parameter, etc., indicate defective regions in the crystal but fail to describe the defect morphology and concentration. There are newer geometrical feature vectors designed with a focus on supervised learning to identify defects from lattice atoms and visualize them \cite{von2020favad, goryaeva2020reinforcing}. The accuracy of machine learning that uses the traditional, hand-crafted approach to feature designing is limited by the relevant information captured in the feature. The advantage of supervised learning for the tasks of defect identification and characterization is not yet clear, especially when the crystal structure and defect morphology have well-defined descriptions amenable to efficient deterministic methods. However, machine learning explorations can guide establishing deterministic definitions. An unsupervised learning method \cite{BHARDWAJ2020109364} used for defect classification performs well in differentiating edge dislocations, C15 like rings, etc. and outlines all the possible defect shapes observed in Fe and W cascades. However, it fails to distinguish mixed clusters and provide details of morphologies such as orientations of dislocation loops and dumbbells constituting a defect.

Several studies indicate the presence of complex morphology in radiation-
induced defects. An early MD study of clusters formed in energetic cascades in W by Sand
et al. \cite{Sand_2013} uses W-S and individual inspection guided by potential energy analysis to report dislocation loops. The study also reports some of the
clusters having complex configurations that often have partial parallel oriented
dumbbells. A subsequent MD study of radiation damage in W across different temperatures and energy range by Wahyu Setyawan et al. \cite{SETYAWAN2015329} presents a detailed
categorization of all SIA clusters using a combination of W-S, ES, and SIA
dumbbell/crowdion orientations. However, it groups SIA clusters of less than size 30 and significant sized clusters that do not form a dislocation loop under the ambiguous category of 3D clusters, which is subcategorized based on the constituent SIA orientations. A study on the effect of repulsive part of interatomic potential on cascade damage in Fe \cite{BYGGMASTAR2018530} uses similar methods. Besides, it labels non-dislocation clusters with dumbbells in specific orientations as C15, which are of much interest due to their sessile nature and high stability in Fe. The method used to label C15 is verified by visual inspection of positive cases found with the technique. Since the structure of the primary radiation damage has significant impact on the mobility, interaction, and annealing behavior of the defects, longer time scale predictions of the microstructural evolution would benefit from a more detailed description of each defect. Furthermore, a recent study by Mason et al. reports the formation of intermediate complex configurations \cite{Mason2019}, as initially separate defects meet and merge during the process of annealing. The identification of the defect morphology at these stages of annealing can potentially provide insights and aid predictions of the time scale and final product of the transformation.

We present a method to define the morphology of a cluster based on its
homogeneous constituent components. A separate component in a
cluster is defined as composed of SIA dumbbells and crowdions that all hold a specific relationship with their neighbors. We represent a defect as a computational graph, with SIA dumbbells/crowdions as nodes and the relationship between them as edges. The problem of identifying the morphology is reduced to finding connected components and cycles in the graph. The method gives structural details of defects such as the overall orientation of a dislocation, number and proportions of different morphologies in a multi-component defect, and degree of disorientation and extent of individual SIAs that constitute a defect. We explore the defect morphologies of over a thousand SIA clusters formed in 149 high energy W collision cascades. We ascertain the presence of 3D rings corresponding to the C15 Laves phase in W and show new insights about other morphologies.

\section{Results}
\label{sec:res}

The dataset used for the analysis has collision cascades in bulk W simulated with an initial temperature at 0 Kelvin and evolved for 40ps. Electronic stopping is applied to atoms with energies above 10 eV \cite{SAN15}. The database contains a total of 139 cascades at 50 keV, 100 keV, 150 keV, and 200 keV simulated with the Derlet potential \cite{PhysRevB.76.054107} stiffened by Bjorkas \cite{BJORKAS20093204}. The cascades contain 1170 clusters of different sizes and morphologies. The defect morphology identification takes approximately a minute to process the database on a regular desktop computer once defects and clusters have been identified. The defect and cluster identification using efficient implementations of W-S and ES employing modular arithmetic  \cite{BHARDWAJ2020109364} takes less than ten minutes for the whole database. In comparison, the dislocation analysis with DXA algorithm \cite{Stukowski_2012} as currently implemented in Ovito \cite{ovito} takes more than two minutes to process a single 150 keV cascade on the same desktop computer, while a 200 keV cascade having a simulation size of 190 unit cells could not be processed due to insufficient system memory.

For identifying the defect morphology, we first find line equations for each dumbbell and crowdion's string of displaced atoms. These lines correspond to the nodes in the graph representation of a defect. Whether to join two nodes in the graph with an edge is predicated upon rules defined separately for parallel and non-parallel ring-like configurations. For a parallel component, a connecting edge between two neighboring nodes is added if the lines they represent are parallel. In contrast, for a ring, an edge is added if the angle between 1NN neighboring lines is approximately 60 or 90 degrees. The connected components algorithm from graph theory is used \cite{tarjan1972connected} to find distinct homogeneous components. To further verify ring morphology, the graph representation must also exhibit cycles of length three or more.
 This verification step adds robustness to the algorithm against errors that can arise due to thermal vibrations. A parallel component forms an edge dislocation by introducing an extra plane of atoms in the crystal. The direction of the Burgers vector is the same as the direction of the lines formed of the strings of displaced atoms, and the number of extra atoms in lines decides the magnitude of the Burgers vector.
The morphologies that are neither parallel nor form rings are transient non-specific configurations. These are meta-stable configurations that quickly change to a glissile parallel bundle of crowdions when annealed at room temperature or below.

The characterization criterion used here is described in detail in the methods section. The motivation for the conditions used for introducing edge connections becomes intuitively clear upon closer inspection of of the defect morphologies, presented in the following results.

\subsection{Defect Morphologies in W collision cascades}

\begin{figure}[H]
  \centerline{\includegraphics[width=1.0\linewidth]{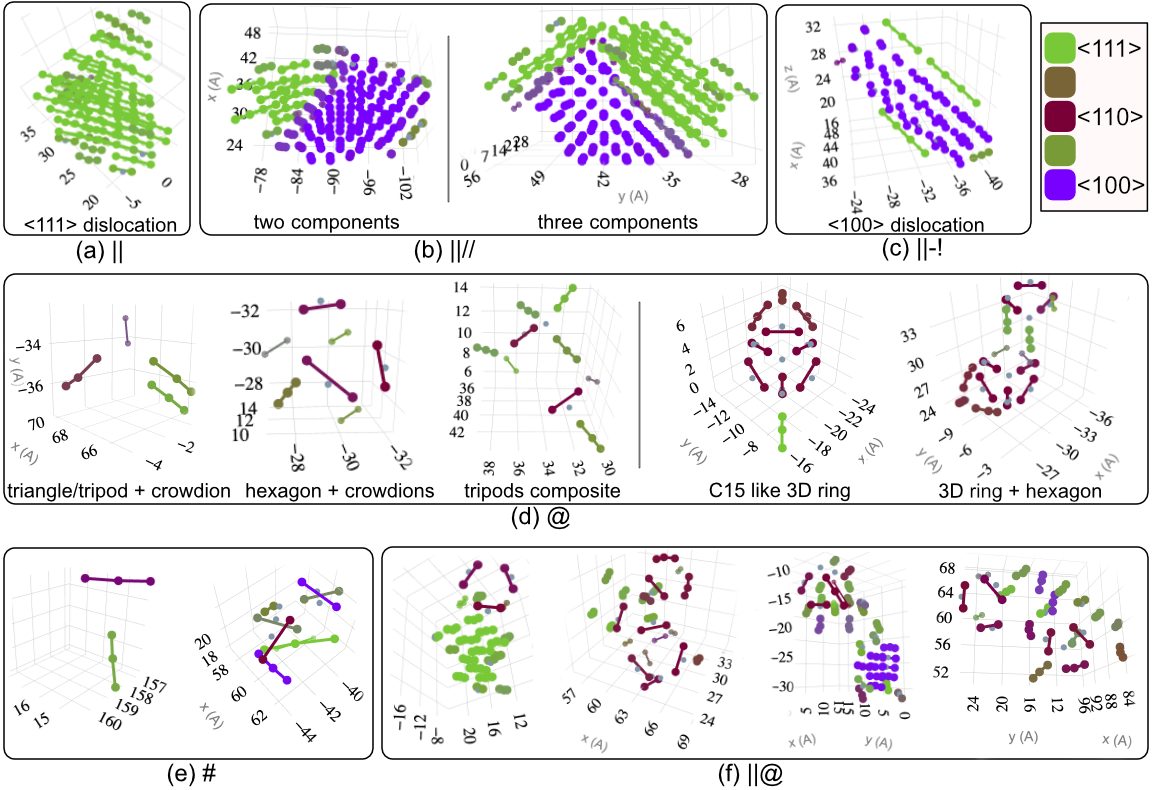}}
  \caption{\label{fig:fig-1}
  Different defect morphology in W collision cascades with the symbols used to represent them. (a) parallel bundle of \3 directed SIA that form 1/2\3 edge dislocations, (b) defects composed of multiple edge dislocations, (c) parallel group of \1 directed SIA that form \1 edge dislocations, (d) 3D rings and their planar ring basis, (e) meta-stable defects with no particular order, (f) defects with rings and edge-dislocations. The first two defects in (d) show planar rings that are basic shapes for 3D rings. Lines are drawn along the dumbbells/crowdions and are colored according to their orientation.
  }
\end{figure}

The morphology of SIA defects in the database of W collision cascades is found to be composed of parallel bundles of SIA strings forming edge dislocations, 3D rings corresponding to C15 structure, their planar basis shapes (triangle/tripod and hexagon \cite{DEZERALD2014219, PhysRevLett.108.025501, BHARDWAJ2020109364}), and a few transient meta-stable configurations. A defect morphology can also be a composition of multiple homogenous components, such as a defect composed of multiple edge dislocations or a defect consisting of a ring and an edge dislocation. \Cref{fig:fig-1} shows typical clusters for each defect morphology. The edge dislocations having \1 orientation are observed to have a few \3 crowdions on the fringes. The direction of dumbbells in 3D rings is mostly \2, while for their planar basis, it can be anywhere between \2 and \3. The rings may have a few \3 crowdions extending as tails. Slight disorientation from \2 direction can also occur if the ring is adjoined with an edge dislocation forming a composite cluster. The planar ring with tripod or triangle di-interstitial form can also compose into structures other than the 3D C15 like rings, such as two of the tripods stacked one after the other. Among the less stable, random arrangements of dumbbells and crowdions, we find a specific configuration of size two cluster that re-occurs, unlike different random transient defect configurations. The cluster is composed of two orthogonal dumbbells that are 2NN distance apart (the first defect in \Cref{fig:fig-1} (e)).

\begin{figure}[H]
  \centerline{\includegraphics[width=1.0\linewidth]{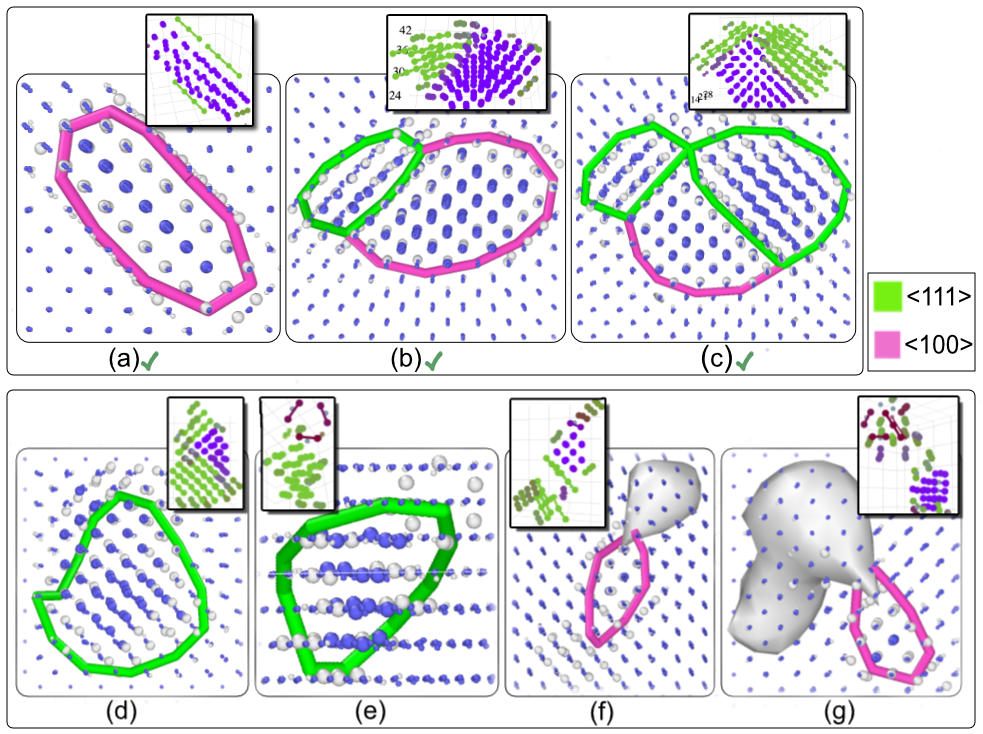}}
  \caption{\label{fig:fig-2}
Comparison of dislocation loops found using dislocation analysis. The first row shows the cases where dislocation analysis agrees with the current method. The dislocation loops in (d) and (e) only show a single \3 loop and miss the other smaller components that include both edge dislocations and hexagonal ring. In (f) and (g), the dislocation represents the other parallel group of SIAs and ring components by connected blobs. The blobs represent non-dislocation defects, while the current method provides specific details for such defects.
}
\end{figure}

The composite morphologies (\Cref{fig:fig-1} (b) and (f)) render different behavior from the individual constituent components, e.g., the movement of an otherwise glissile dislocation loop can be restricted by being trapped with a ring or in conjunction with another dislocation loop. \Cref{fig:fig-2} shows some cases where dislocation analysis results (found with Ovito\cite{ovito}) alone can result in the omission of such distinctions in composite clusters.

\subsection{Statistical distribution of defect morphology}

\begin{figure}[H]
  \centerline{\includegraphics[width=1.0\linewidth]{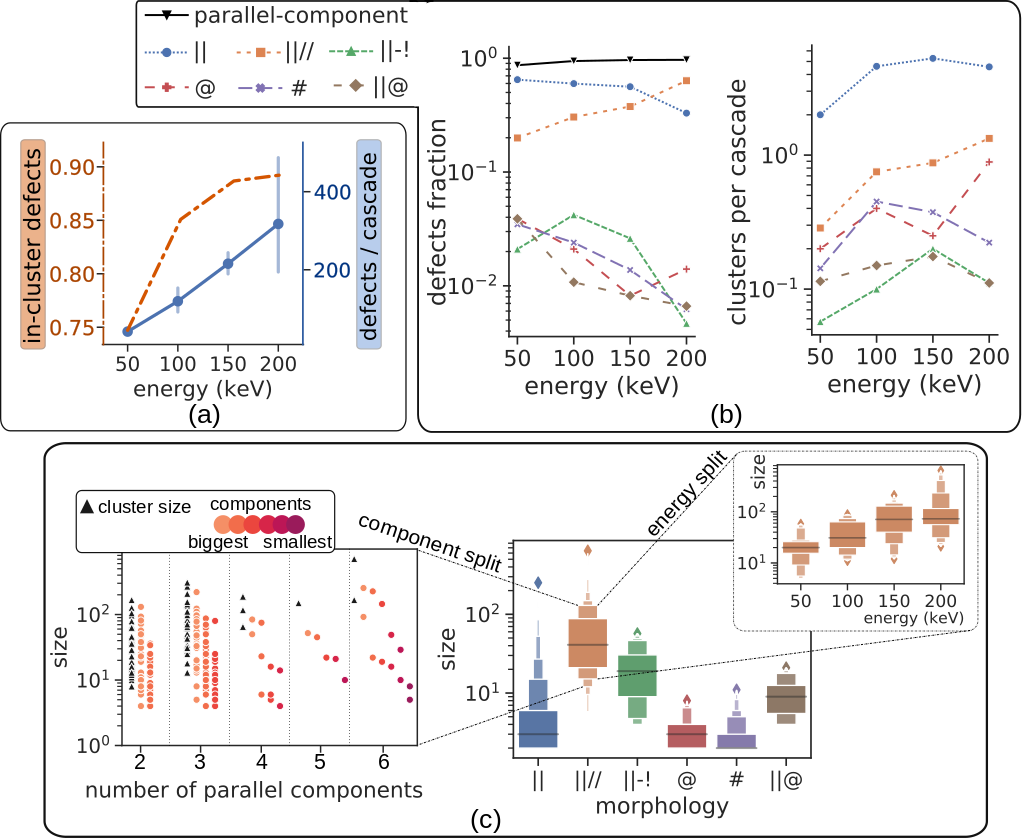}}
  \caption{\label{fig:fig-3}
Statistics of defects formed in W collision cascades. (a) Shows an average number of point defects per cascade and fraction of point defects in a cluster. (b) Shows the relative distribution of in-cluster defects among morphologies across PKA energies. The second plot shows the average number of times a defect morphology appears in a cascade. (c) Shows the size distribution of each morphology. The size for multi-component loops (||//) shows a definite increasing trend with energy and is shown separately on the right inset. Other morphologies do not show a clear trend with energy. The number of components and their sizes are shown on the left.
  }
\end{figure}

\Cref{fig:fig-3} shows the statistics of defect concentration over different energies and the distribution of the defect morphologies for the complete dataset of 1173 point defect clusters. The majority of defects arrange in parallel bundles of dumbbells and crowdions, especially in smaller, glissile \3 orientations and big mixed, sessile composites of \3 - \1 multi-component dislocation loops. \Cref{fig:fig-3} (b) shows the fraction of defects forming a parallel bundle of dumbbells/crowdions (black inverted triangle), which includes \3 oriented bundles (blue circles), \1 bundles (green triangle) and combinations of \3 - \1 (orange square). The number of the multi-component edge dislocations and their sizes (right inset in \Cref{fig:fig-3} (c)) increase with energy. The number of distinct loops in a single multi-component dislocation goes up to six (left inset in \Cref{fig:fig-3} (c)).

The fraction of point defects in non-parallel defect morphologies decreases with energy while the number of defects per cascade increase (\Cref{fig:fig-3} (b)), implying a decrease in the sizes of these defects. Compared to dislocations, these defects are fewer; however, their role in the evolution of a cascade might be significant depending on their thermal stability and interactions.

\subsection{Exploring internal morphplogical details}

The method gives insights into internal details of defect morphologies. For example, it has been postulated that the strings of displaced atoms in the central part of a 1/2\3 dislocation are more extended, forming longer crowdions than the SIA lines on the surface\cite{Dudarev111clusters2003}. The number of SIA lines within 1NN is a good indicator of whether an SIA line is on the cluster's surface or towards the center in bulk. For a \3 parallel cluster, the maximum number of neighbors for a central SIA line is six, while for \1, this value is four.

\begin{figure}[H]
  \centerline{\includegraphics[width=.5\linewidth]{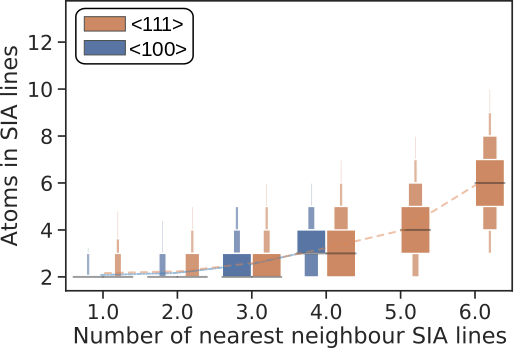}}
  \caption{\label{fig:fig-4}
  Number of atoms in an SIA line as a function of its nearest neighboring
  SIAs. The number of neighbors indicates whether an SIA is in the central part of the cluster or on the surface. The plot shows that the extent of an SIA line in the central part having six neighbors is longer and is almost always more than the base value of two. For \1 SIAs, the central SIA lines with four neighbors become longer on average, but a good fraction has only two atoms. The extent of SIA lines in \1 rarely goes beyond three atoms, while \3 crowdions are longer.
  }
\end{figure}

We find only dumbbells (SIA lines with only two displaced atoms) in the \2 direction. The extents of displaced atoms in \1 rarely go beyond three atoms while in \3 longer strings of displaced atoms form (see \Cref{fig:fig-4}). For \3 oriented dislocations, there are longer strings of displaced atoms in the central SIA lines. The correlation value between the extent of an SIA line and its number of neighbors is 0.78. While for \2, the correlation value is just 0.44.


\section{Discussion\label{sec:discuss}}
The presented approach identifies defect morphologies and also resolves constituent uniform morphologies in a multi-component defect. A morphologically homogenous component is composed of SIA dumbbells/crowdions that share a characteristic neighborhood relationship. The reduction of the problem to well-established problems from graph theory viz. connected components and cycles in a graph make the solution efficient and easy to extend to a new morphology by defining its fundamental characteristics as a decision rule for connecting graph nodes. The implementation of the algorithm is fast, making it ideal for use in big datasets.

For the identification of dislocations, the method is validated by comparing it to the results of dislocation analysis. The presented method can give a more accurate description of multi-component dislocations and mixed ring and dislocation defects. We show examples for cluster morphologies where dislocation analysis alone may not be sufficient to predict cluster structure and its implied behavior accurately. 

The method is applied to a dataset of 149 high energy collision cascades in W consisting of over a thousand SIA defect clusters. The analysis identifies the morphology of every SIA defect cluster. The in-cascade defects in W show different morphologies such as single dislocations along \3 and \1 orientations and their multi-component composites, 3D rings corresponding to the C15 Laves phase, and its planar basis rings. The method also identifies mixed defects composed of dislocations and rings. The statistics on fraction and defect size of different morphologies show that glissile \3 edge dislocations of small size are predominant. However, as the PKA energy increases, the number and size of sessile multi-component dislocations increase. Although rarer than dislocations, the sessile rings can play an essential role in the cascade's thermal evolution by acting as trapping sites for dislocations due to their sessile nature. However, the extent of their role will depend on the morphological stability and nature of interactions with other defects. The distribution of in-cascade defect morphology is an essential metric for validation and comparison of interatomic potentials and treatment of electronic stopping used in the MD simulations. While recent experiments have validated the presence of \1 dislocations in W \cite{yi2012, Yi_2015}, in addition to 1/2\3 dislocations, the experimental investigation of smaller defect morphologies found in simulations is still intractable with conventional transmission electron microscopy\cite{andreaMason2017}. Higher scale models that use the defect morphology distribution and their properties can help in such validations.

The presented method provides structural details such as the orientation and extent of individual SIA string and its relationship with neighbors. We use the information to quantify the postulated correlation \cite{Dudarev111clusters2003} between the extent of an SIA line and whether it is in the central part of the cluster. The details can be used to study the mechanisms of morphological transitions and migration of defects, providing insights on the relationship between structural details, size of the defect, and defect properties like transition energy, recombination radius, and migration energy. These can be input to higher scale models such as Monte Carlo methods, which currently use approximate functions for such essential relationships \cite{lakimoca}. These higher scale predictions of the microstructure can be validated with the current experiments.

Identifying a defect morphology requires defining basic rules that are characteristic of that morphology. To identify a new defect morphology, its basic structure must be understood, apriori, which can be a weakness of the method if many unknown or hard to define morphologies exist in the application domain. However, in the domain of crystals, the defects are generally limited in number and have definite structures. Another limitation of the method arises from the thermal noise inherent in dynamic simulations at finite temperatures. While a reasonable degree of thermal fluctuations poses no problems to W-S or ES analysis, at the level of sensitivity of the method presented here, these transient displacements may rarely induce errors, potentially changing the angle of a line relative to its neighbors. In a large defect, these spurious errors do not affect the overall morphology, however small defects with noise may appear as random arrangements. This issue can be trivially circumvented by energy minimization of the defect structures, but for large databases, such a step would entail a significant computational cost. The extra check for cycles in the graph representing a probable ring component increases robustness against misclassification to rings due to spurious thermal noise while also incurring a low computational cost. 

\section{Methods}
\label{sec:impl}

\begin{figure}[ht]
  \centerline{\includegraphics[width=.9\linewidth]{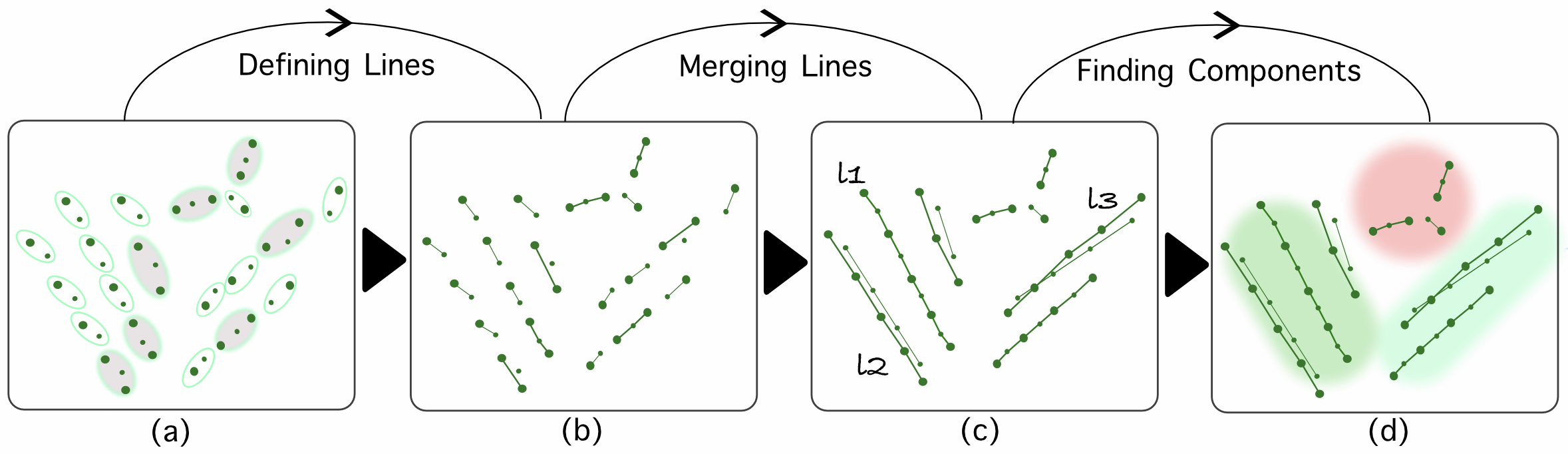}}
  \caption{\label{fig:fig-5}
    The figure schematically shows the output for a cluster after each step.
    (a) Initial inputs from the defect identification: coordinates of the displaced atoms along with the nearest lattice site for each atom. (b) Define lines by (i) joining the pair of atoms that occupy the same lattice site, (ii) displaced atom and lattice site if only a single atom occupies the lattice site. (c) Merge the coinciding lines. (d) Find structurally homogeneous components that constitute the defect based on angle and distance relationship between the neighboring lines. 
    }
\end{figure}

An efficient implementation of W-S and ES based method that employs modular arithmetic to associate each atom to its nearest lattice site \cite{BHARDWAJ2020109364} is used to find the point defect clusters from given atomic coordinates at a time step. The displaced atom coordinates of a defect and their nearest lattice points constitute the initial input of the morphology identification. \Cref{fig:fig-5} summarizes the steps in the method consisting of (i) defining lines along the displaced atoms and associated lattice points, (ii) merging coinciding lines, and (iii) finding connected components in the graph representation of a defect.

\subsection{Defining and Merging Lines}
\label{sec:lineimpl}
\begin{enumerate}
  \item Define parametric line equations for lattice sites associated with a pair of atoms as the line passing through the two atoms. The lattice site itself may not lie on the line. For the lattice site associated with only a single displaced atom, the line is defined as passing through the lattice site and the displaced atom (Fig. 2(b)).
  \item Find and merge neighboring coincident lines and collinear points (Fig. 2(c)). This operation can employ k-d tree data-structure \cite{kdtree} to look for neighboring lines efficiently.
\end{enumerate}
 
In Fig. 2(c), a single line L1 is defined by merging the coincident lines. As shown by L2 and L3 lines in Fig. 2(c), for some crowdions, the lattice points can be seen as falling in a separate sub-line, possibly due to local stresses caused by nearby non-parallel structures or other clusters. When visualized, the lines help in the qualitative assessment of different structures such as parallel crowdions, hexagonal rings, 3D-rings, and mixed clusters.

We use parametric equations of lines for efficiently calculating different
properties such as the shortest distance between two lines, angle between two lines and orientation of a line. For the parametric equation of a line passing through two points a and b, we define a unit direction vector $\vec v = \vec a - \vec b / \|\vec a- \vec b\| $. The equations for angle between two lines ($\theta$) and their shortest distance ($d$) can be defined as:

\begin{equation}
\label{eq:angle}
 \theta = \arccos{\langle \vec v_1, \vec v_2\rangle}
\end{equation}

\begin{equation}
\label{eq:lineDist}
d = \frac{\langle \vec a_1-\vec a_2,\vec v_1\times \vec v_2\rangle}{\|\vec v_1\times \vec v_2\|}
\end{equation}

where $\vec v_1$ and $\vec v_2$ are direction vectors of the lines and $\vec a_1$ and $\vec a_2$ are any two points in the lines. For coincident lines, $\theta$ and $d$ both should be close to zero. We will use these metrics for graph representation also.

The unit direction vector also represents the orientation of the line. We also add extra line attributes like defect count for each line, their extent, deviation from the standard orientations, and offset of lattice sites that do not lie on the line. The defect counts for the constituent lines in an edge dislocation indicate the magnitude of the Burgers vector for the dislocation.

\subsection{Graph Representation}
\label{sec:graphimpl}

After merging the coinciding lines, we construct the adjacency matrix $A$, which is an $n \times n$ sized matrix for a cluster with $n$ number of lines. Each value $a_{ij}$ of the matrix is either $1$ if the $i^{th}$ and $j^{th}$ lines are marked as connected or $0$ if they are not connected. The connectivity rules are defined differently for bundles of parallel dumbbells and planar rings or C15-like 3D-rings.

\subsubsection{Edge Predicate for Parallel component}
\label{sec:parallelimpl}

For components consisting of parallel dumbbells/crowdions the adjacency matrix values are defined by the following relation:

\begin{equation}
\label{eq:parallelComp}
  a_{i,j}=
    \begin{cases}
      1, & \text{if}\ \theta \approx 0 \ \text{and}\ d \leq 1NN \\
      0, & \text{otherwise}
    \end{cases}
\end{equation}

$\theta$ is the angle between the lines, and $d$ is the shortest distance between the two lines found using the \Cref{eq:lineDist}.

\subsubsection{Edge Predicate for Ring component}
\label{sec:ringimpl}

The values for adjacency matrix of rings are defined by the following relation:

\begin{equation}
\label{eq:ringComp}
  a_{i,j}=
    \begin{cases}
      1, & \text{if}\ \theta \approx 60 \ \text{or}\ 90 \ \text{and}\ d^{\prime} = 1NN \ \text{or}\ 3NN \\
      0, & \text{otherwise}
    \end{cases}
\end{equation}

$\theta$ is the angle between the lines, and $d^{\prime}$ is the shortest distance between any lattice points associated with the line. Both 60$^\circ$ or 90$^\circ$ and 1NN or 3NN are valid values for different time snapshots of the di-interstitial tripod-like arrangement, while for tri-interstitial hexagon, 3NN and 60$^\circ$ suffice.

\begin{figure}[H]
  \centerline{\includegraphics[width=.8\linewidth]{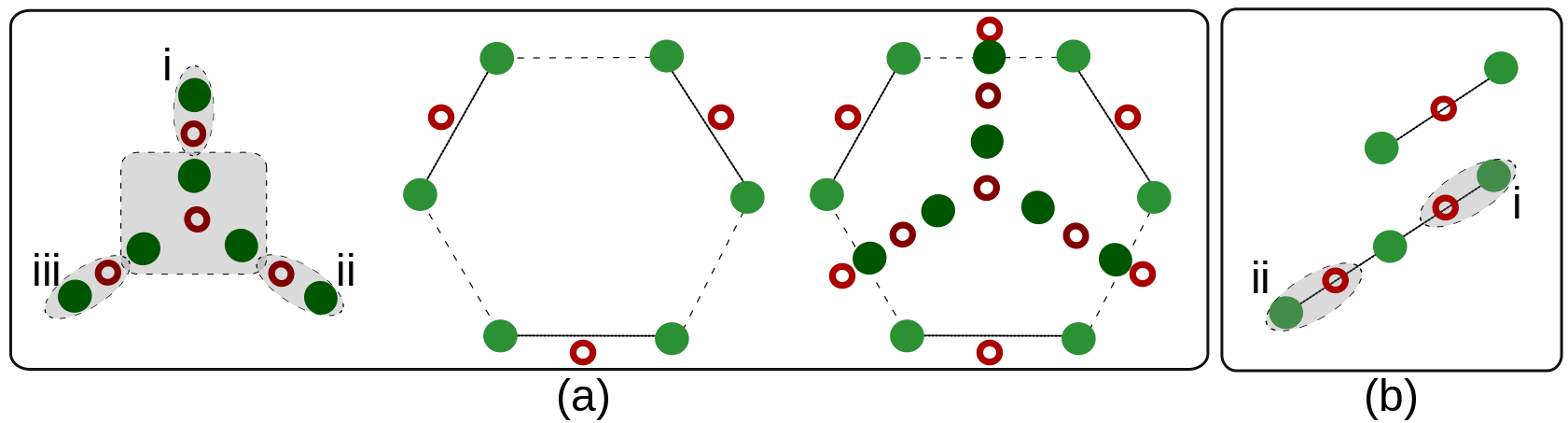}}
  \caption{\label{fig:fig-6}
The figure shows a schematic representation of basic cluster shapes and the effect of thermal vibrations on their different possible appearances. Green circles represent interstitials, and red hollow circles represent lattice points. (a) shows the tripod/triangle and hexagon as a basis that forms the 3D ring shape. The triangular di-interstitial appears in four forms depending on how many extra pairs of displaced atoms and corresponding lattice sites (labeled with i, ii, iii) appear in a particular time-step. If all the three extra pairs appear displaced, then it appears as a tripod with a vacancy at the center. The non-planar tripod appears planar in the schematic. (b) shows that one of the dumbbells in a size two parallel cluster may appear either as a crowdion or dumbbell depending on the atomic vibration at the time-step. 
}
\end{figure}

A 3D-ring or C15-like structure can be defined as a composite of di-interstitial tripod structure and tri-interstitial hexagonal shape \cite{DEZERALD2014219, PhysRevLett.108.025501} (\Cref{fig:fig-6} (a)). Both of these structures are stable by themselves and also occur as a stand-alone sessile cluster. To understand the adjacency rules defined in \Cref{eq:ringComp}, we require the structural details of these two basic shapes.

The hexagonal structure is a planar arrangement of three dumbbells whose lattice sites are not collinear. The lines drawn by joining the dumbbell atoms appear as alternate sides of a hexagon, forming a sixty-degree angle with each other. All the three lines are oriented in \2 direction, and the lattice points of lines are 3NN distance apart. The clusters that have this arrangement also have another crowdion/dumbbell that is orthogonal to the plane of the hexagon and is oriented along \3 direction.

The di-interstitial tripod, when appearing as a stand-alone cluster, can appear in four different forms in different time snapshots (\Cref{fig:fig-6}(a)) due to thermal vibrations.  These four forms and their geometries can be observed if we look at multiple time-instances of an MD simulation of this defect morphology. For all the forms, angles are either ninety or sixty degrees, distances between lattice sites are either 1NN or 3NN, and line orientations are between \2 and \3. A tripod can also sometimes be augmented with a single or a couple of crowdions/dumbbells. 

\subsection{Finding the components}

Using the connected components algorithm, we group all the nodes (representing SIA lines) that are connected by the same type of edges (defined in \Cref{eq:parallelComp} and \Cref{eq:ringComp}). A parallel component forms an edge dislocation with the Burgers vector having direction along the constituent lines and magnitude determined by the number of extra atoms in the lines. After we have found parallel components and ring-like components, a small fraction of the remaining components are found to be SIAs arranged in no specific configuration. These are mostly non-recurring transient configurations. However, one specific meta-stable configuration that appears more than once is an orthogonal pair of 2NN separated dumbbells (shown in \Cref{fig:fig-1} (e)). It, too, like any other non-specific configuration, quickly changes to glissile \3 parallel form at around room temperature.

\subsection{Check for cycles in ring components}
\label{sec:cycleimpl}

For a ring structure, in addition to the binary relation defined by the \Cref{eq:ringComp}, it is also essential that the three lines in a hexagon, tripod, or their 3D composite mutually hold the same relationship. More specifically, if $a_{ij} = 1$ and $a_{jk} = 1$, $i, j, k$ will be grouped together into one component, however for a ring it is also essential that $a_{ik} = 1$ otherwise it is a random non-parallel arrangement of dumbbells and not a ring. These are called triangle graphs or C3 (cycles of size 3) in a graph \cite{triangleGraph}. After finding the connected components, we look for these cycles of three to verify a ring-like structure. There are also bigger cycles in a 3D-ring, and the biggest cycle contains the complete ring shape. Another way of differentiating between planar and 3D rings is that in a 3D-ring there are multiple C3 cycles, while in a planar ring (tripod or hexagonal ring), there is only a single cycle as there are only three lines that hold the relationship defined in \Cref{eq:ringComp}. The probable ring components that do not form such cycles are classified as random transient configurations that are neither parallel nor ring. This condition takes care of spurious thermal noise that may appear at higher temperatures.

\bibliographystyle{naturemag}

\end{document}